\documentclass{nature}

\usepackage{xcolor}
\usepackage{dcolumn}
\usepackage{bm}

\usepackage[
margin=0.5in,
]{geometry}

\usepackage{graphicx}
\usepackage{multicol}
\makeatletter
\let\saved@includegraphics\includegraphics
\AtBeginDocument{\let\includegraphics\saved@includegraphics}
\renewenvironment*{figure}{\@float{figure}}{\end@float}
\makeatother
\usepackage{subcaption}
\usepackage{enumerate}
\usepackage[shortlabels]{enumitem}
\usepackage{amssymb}
\usepackage{amsthm}
\usepackage{amsmath,scalerel}

\usepackage{hyperref}

\usepackage{braket}
\let\baraccent=\= 
\renewcommand{\=}[1]{\stackrel{#1}{=}} 

\newtheoremstyle{mystyle}
  {}                                      
  {}                                      
  {}
  {}                                      
  {\itshape}
  {:}                                     
  { }                                     
  {\thmname{#1}\thmnumber{ #2}\thmnote{ (#3)}}%

\theoremstyle{mystyle}
\newtheorem{theorem}{Theorem}

\makeatletter
\newcommand{\setword}[2]{%
  \phantomsection
  #1\def\@currentlabel{\unexpanded{#1}}\label{#2}%
}
\makeatother

\newcommand{\nth}{\bar{n}_{\scriptsize\textrm{th}}}

\title{Enhanced energy-constrained quantum communication over bosonic Gaussian channels}

\author{Kyungjoo Noh$^{1,2}$, Stefano Pirandola$^{3,4}$, Liang Jiang$^{1,2,5}$}

\begin{document}

\maketitle

\begin{affiliations}
 \item Departments of Applied Physics and Physics, Yale University, New Haven,
Connecticut 06511, USA
 \item Yale Quantum Institute, Yale University, New Haven, Connecticut 06520, USA
 \item Computer Science and York Centre for Quantum Technologies, University of York, York YO10 5GH, UK
 \item Research Laboratory of Electronics, Massachusetts Institute of Technology (MIT), Cambridge, Massachusetts 02139, USA
 \item Pritzker School of Molecular Engineering, University of Chicago, 5640 South Ellis Avenue, Chicago, Illinois 60637, USA 
\end{affiliations}

\begin{abstract}
Abstract 

Quantum communication is an important branch of quantum information science, promising unconditional security to classical communication and providing the building block of a future large-scale quantum network. Noise in realistic quantum communication channels imposes fundamental limits on the communication rates of various quantum communication tasks. It is therefore crucial to identify or bound the quantum capacities of a quantum channel. Here, we consider Gaussian channels that model energy loss and thermal noise errors in realistic optical and microwave communication channels and study their various quantum capacities in the energy-constrained scenario. We provide improved lower bounds to various energy-constrained quantum capacities of these fundamental channels and show that higher communication rates can be attained than previously believed. Specifically, we show that one can boost the transmission rates of quantum information and private classical information by using a correlated multi-mode thermal state instead of the single-mode thermal state of the same energy. 

\end{abstract}

\newpage

\begin{multicols}{2}
[]

\section*{Introduction}



Quantum communication is a field of quantum information science that takes advantage of unique quantum mechanical nature of information carriers to realize secure classical communications\cite{Bennett1984,Pirandola2019} and build a large-scale quantum network\cite{Kimble2008,Pirandola2016,Wehner2018}. Realistic quantum communication channels are noisy and therefore quantum error correction or entanglement distillation\cite{Bennett1996a,Bennett1996b,Bennett1996c} is essential to faithfully implement various quantum communication tasks. Due to the resource overhead associated with the error correction or entanglement distillation, quantum communication rates are fundamentally limited by the noise in quantum channels that are used to transmit quantum information. Thus, determining the maximum achievable quantum communication rates of an experimentally relevant noisy quantum channel is of fundamental importance to the quantum information science.

A celebrated result due to C. E. Shannon\cite{Shannon1948}
established that the maximum achievable (classical) communication
rate of a (classical) channel equals the channel's mutual
information, or the channel's (classical) capacity. The notion of
(classical) channel capacity has been generalized to the quantum
realm\cite{Nielsen2000,Hayashi2016,Watrous2018}, and there are
various notions of quantum capacities that characterize the
channel's maximum achievable quantum communication rates for
various quantum communication tasks (see, e.g., Ref.
\citen{Watrous2018} Sec. 8). For example, the regularized private
information $P_{\textrm{reg}}(\mathcal{N})$ of a channel
$\mathcal{N}$ quantifies the maximum achievable secure classical
communication rate of the channel, also known as the channel's
private capacity\cite{Devetak2005}. Similarly, the regularized
coherent information $Q_{\textrm{reg}}(\mathcal{N})$ quantifies
the maximum achievable quantum state transmission rate (without
classical feedback assistance), also known as the channel's
quantum
capacity\cite{Schumacher1996,Lloyd1997,Barnum1998,Barnum2000,Devetak2005}.

Evaluation of these quantities however requires optimization over
all input states to infinitely many channels\cite{DiVincenzo1998,Smith2008,Smith2011,Cubitt2015,Lim2018,Lim2019}
and therefore is generally intractable\cite{Wolf2011,Oskouei2018}
unless the channel has a special structure such as degradability
or anti-degradability\cite{Devetak2005C,Caruso2006,Yard2008}.
Moreover, explicit formulas for the two-way quantum capacity
(i.e., maximum quantum state transmission rate with two-way
classical feedback assistance) are only known for so-called
distillable channels\cite{Pirandola2017}
i.e., channels whose Choi matrices have relative
entropy of entanglement\cite{Vedral1997,Vedral1998,Vedral2002} equal to the one-way distillable
entanglement. Even though the two-way quantum (and private)
capacities of these channels are known\cite{Pirandola2017} and
also have suitable generalizations to repeater chains and quantum
networks of arbitrary
topology\cite{Pirandola2019b,Pirandola2019m}, finding similar
results for other types of quantum channels is the subject of
intensive investigation, especially within the setting of
continuous variable systems\cite{Braunstein2005}.


In this work, we consider bosonic Gaussian
channels\cite{Weedbrook2012} and study their various quantum
capacities in the energy-constrained
scenario\cite{Giovannetti2003}. Among Gaussian channels,
thermal-loss channels characterize energy loss and
thermal noise errors and thus model realistic optical and
microwave quantum communication channels, i.e., two leading platforms for quantum communication technologies.
Therefore, understanding the quantum capacities of Gaussian
channels is of great practical importance as well as of academic
interest. Previously, it was shown that bosonic pure-loss channels
(a subclass of thermal-loss channels that do not
have thermal noise) are either degradable or
anti-degradable\cite{Caruso2006}. Thus, various quantum capacities
of these subclass of Gaussian channels are well understood and
determined analytically\cite{Holevo2001,Wolf2007,Pirandola2017}.

In practice, there is additional thermal noise
added to the communication channels, which can be induced by the
laser noise in optical communication or by the background thermal
noise in microwave communication\cite{Xiang2017,Axline2018}. Hence, it is
important to understand the quantum channel capacities for these
more general thermal-loss channels. However, thermal-loss
channels are neither degradable nor
anti-degradable\cite{Caruso2006B,Holevo2007} and only
lower\cite{Holevo2001,Carcia2009,Pirandola2009,Ottaviani2016,Sharma2018}
and upper
bounds\cite{Holevo2001,Pirandola2017,Sharma2018,Rosati2018,Noh2019}
are known for their various quantum and private capacities
(one-way, two-way, energy-constrained or unconstrained).

In this paper, we establish improved lower bounds of various energy-constrained quantum capacities of thermal-loss channels that are stronger than the existing bounds. That is, we show that higher quantum communication rates can be achieved than previously believed. Specifically, we construct a family of multi-mode Gaussian states, called correlated multi-mode thermal states, and show that they yield larger coherent information (per channel use) than the corresponding single-mode thermal state of the same energy in the low input energy regime. We also show that higher two-way quantum communication rates can be achieved by using correlated multi-mode thermal states and hybridizing forward and backward strategies, instead of using single-mode thermal states and exclusively using a forward or a backward strategy. Finally, we apply a similar technique to further improve the lower bound of the energy-constrained private capacity of the thermal-loss channel.

\section*{Results}

\subsection{Correlated multi-mode thermal states}

\begin{figure}
\centering
\includegraphics[width=7.0cm]{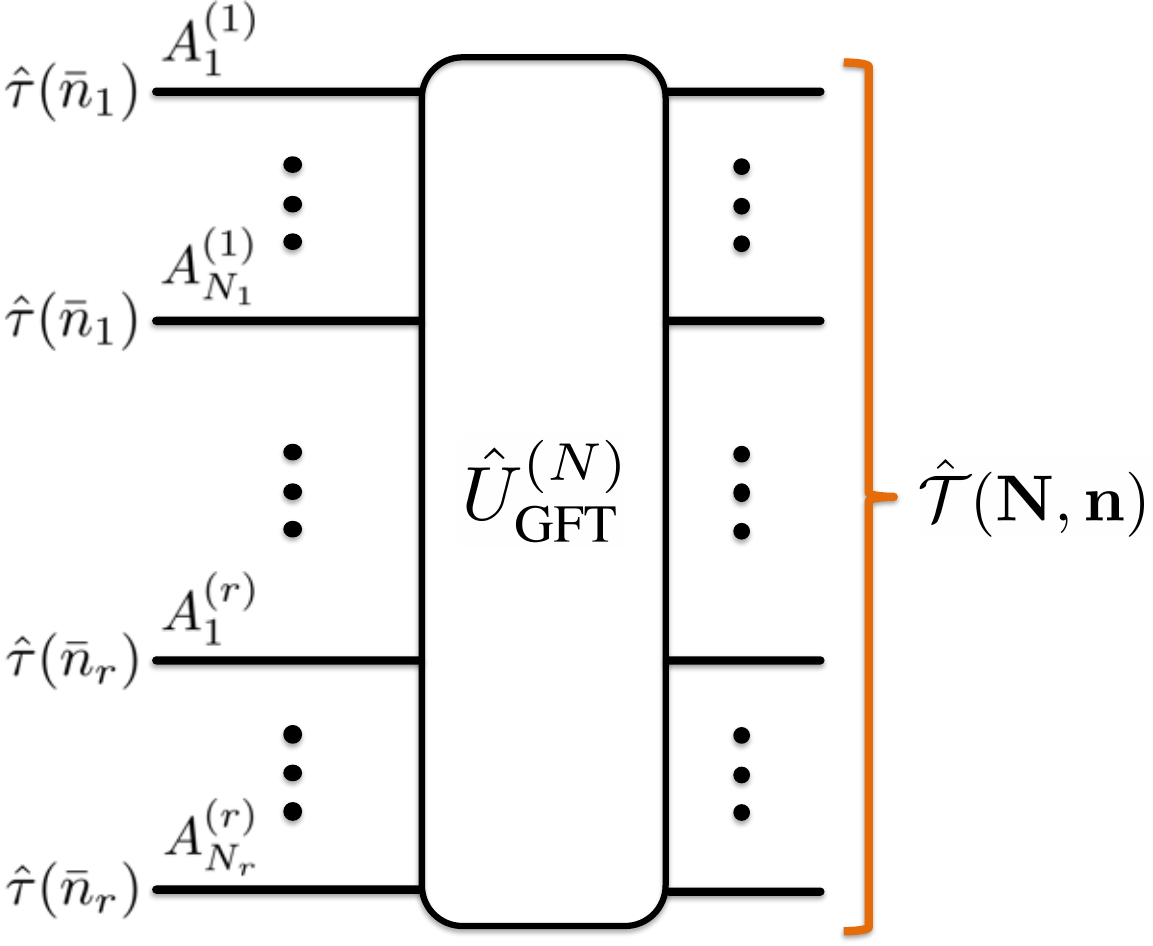}
\caption{\textbf{Generation of a correlated multi-mode thermal state.} A correlated multi-mode thermal state $\hat{\mathcal{T}}(\boldsymbol{\mathrm{N}},\boldsymbol{\mathrm{n}})$ with $\boldsymbol{\mathrm{N}} = (N_{1},\cdots, N_{r})$ and $\boldsymbol{\mathrm{n}} = (\bar{n}_{1},\cdots ,\bar{n}_{r})$ (such that $\sum_{k=1}^{r}N_{k} = N$) can be generated by applying the $N$-mode Gaussian Fourier transformation $\hat{U}_{\textrm{GFT}}^{(N)}$ to an uncorrelated thermal state $\big{\lbrace}\hat{\tau}(\bar{n}_{1})\big{\rbrace}^{\otimes N_{1}}\otimes \cdots \otimes \big{\lbrace}\hat{\tau}(\bar{n}_{r})\big{\rbrace}^{\otimes N_{r}}$.   
}
\label{fig:correlated multi-mode thermal states}
\end{figure}

We first construct a family of Gaussian multi-mode states, called
correlated multi-mode thermal states, which is the key ingredient
for improving the lower bounds of various quantum capacities of
Gaussian channels (see Ref.~\citen{Weedbrook2012} or Methods for the
definition of Gaussian states and channels). Let
$\hat{\tau}(\bar{n})$ denote the single-mode thermal state with an
average photon number $\textrm{Tr}[\hat{n} \hat{\tau}(\bar{n})] =
\bar{n}$, that is $\hat{\tau}(\bar{n}) \equiv
\sum_{n=0}^{\infty}\frac{\bar{n}^{n}}{(1+\bar{n})^{n+1}}|n\rangle\langle
n|$, where $|n\rangle$ is a Fock state. Uncorrelated multi-mode
thermal states would then simply be given by a tensor product of
single-mode thermal states
$\big{\lbrace}\hat{\tau}(\bar{n})\big{\rbrace}^{\otimes N}$. Now
we define correlated multi-mode thermal states as follows:
\begin{align}
\hat{\mathcal{T}}(\boldsymbol{\mathrm{N}},\boldsymbol{\mathrm{n}}) \equiv \hat{U}_{\textrm{GFT}}^{(N)} \Big{[} \big{\lbrace}\hat{\tau}(\bar{n}_{1})\big{\rbrace}^{\otimes N_{1}}\otimes \cdots \otimes \big{\lbrace}\hat{\tau}(\bar{n}_{r})\big{\rbrace}^{\otimes N_{r}} \Big{]} \big{(} \hat{U}_{\textrm{GFT}}^{(N)} \big{)}^{\dagger}.
\end{align}
Here, $\boldsymbol{\mathrm{N}} = (N_{1},\cdots, N_{r})$ such that $\sum_{k=1}^{r}N_{k} = N$ and $\boldsymbol{\mathrm{n}} = (\bar{n}_{1},\cdots ,\bar{n}_{r})$. $\hat{U}_{\textrm{GFT}}^{(N)}$ is the $N$-mode Gaussian Fourier transformation whose action on the $j^{\textrm{th}}$ annihilation operator $\hat{a}_{j}$ is defined as follows
\begin{align}
\big{(} \hat{U}_{\textrm{GFT}}^{(N)} \big{)}^{\dagger} \hat{a}_{j} \hat{U}_{\textrm{GFT}}^{(N)} &= \frac{1}{\sqrt{N}}\sum_{k=1}^{N}e^{i\frac{2\pi}{N}(j-1)(k-1)}\hat{a}_{k} \label{eq:Gaussian Fourier transforamtion}
\end{align}
which holds for all $j\in\lbrace 1,\cdots, N \rbrace$. Hence, the correlated multi-mode thermal state $\hat{\mathcal{T}}(\boldsymbol{\mathrm{N}},\boldsymbol{\mathrm{n}})$ is a collection of single-mode thermal states (where each of the first $N_{1}$ modes supports on average $\bar{n}_{1}$ photons, each of the next $N_{2}$ modes supports on average $\bar{n}_{2}$ photons and so on) which are uniformly mixed by the Gaussian Fourier transformation $\hat{U}_{\textrm{GFT}}^{(N)}$ (see Fig. \ref{fig:correlated multi-mode thermal states}). We remark that each mode in the correlated $N$-mode thermal state $\hat{\mathcal{T}}(\boldsymbol{\mathrm{N}},\boldsymbol{\mathrm{n}})$ supports on average $\bar{n} = \frac{1}{N}\sum_{k=1}^{r} N_{k}\bar{n}_{k}$ photons.

A simple non-trivial example of correlated multi-mode thermal
states would be $\hat{\mathcal{T}}(\boldsymbol{\mathrm{N}},\boldsymbol{\mathrm{n}})$ with $\boldsymbol{\mathrm{N}}
= (1,N-1)$ and $\boldsymbol{\mathrm{n}} = (N\bar{n},0)$ and its covariance
matrix\cite{Weedbrook2012} is given by
\begin{align}
V  = \begin{bmatrix}
(\bar{n}+\frac{1}{2})I_{2}& \bar{n}I_{2} & \cdots & \bar{n}I_{2}\\
\bar{n}I_{2}&(\bar{n}+\frac{1}{2})I_{2}&\cdots& \bar{n}I_{2}\\
\vdots&\vdots&\ddots&\vdots\\
\bar{n}I_{2}&\bar{n}I_{2}&\cdots &(\bar{n}+\frac{1}{2})I_{2}
\end{bmatrix},
\end{align}
where $I_{2}$ is the $2\times 2$ identity matrix. As can be seen from the diagonal elements of the covariance matrix, every mode supports on average $\bar{n}$ photons. Therefore, the reduced density matrix of each mode is given by a single-mode thermal state $\hat{\tau}(\bar{n})$. On the other hand, the off-diagonal elements of the covariance matrix indicate that the position (or the momentum) quadratures of every pair of modes are positively correlated: This is what distinguishes $\hat{\mathcal{T}}(\boldsymbol{\mathrm{N}},\boldsymbol{\mathrm{n}})$ from the uncorrelated $N$-mode thermal state $\big{\lbrace}\hat{\tau}(\bar{n})\big{\rbrace}^{\otimes N}$ and why we call it a correlated multi-mode thermal state.

We remark that correlated multi-mode thermal states can be efficiently prepared because the Gaussian Fourier transformation $\hat{U}_{\textrm{GFT}}^{(N)}$ can be implemented efficiently by using a variant of the fast Fourier transform technique\cite{Cooley1965}.

\subsection{Coherent information and quantum capacity}

Let $\mathcal{N}$ be a quantum channel and $I_{\textrm{c}}(\mathcal{N},\hat{\rho})$ denote the channel's coherent information with respect to an input state $\hat{\rho}$, i.e.,
\begin{align}
I_{\textrm{c}}(\mathcal{N},\hat{\rho}) \equiv S\big{(} \mathcal{N}(\hat{\rho}) \big{)} -  S\big{(} \mathcal{N}^{\textrm{c}}(\hat{\rho}) \big{)},
\end{align}
where $S(\hat{\rho}) \equiv -\textrm{Tr}[\hat{\rho}\log_{2}\hat{\rho}]$ is the quantum von Neumann entropy of a state $\hat{\rho}$ and $\mathcal{N}^{\textrm{c}}$ is the complementary channel of $\mathcal{N}$. The quantum capacity $C_{\textrm{Q}}(\mathcal{N})$ of the channel $\mathcal{N}$ (i.e., the maximum achievable quantum state transmission rate without classical feedback assistance) is equal to the channel's regularized coherent information $Q_{\textrm{reg}}(\mathcal{N})$\cite{Schumacher1996,Lloyd1997,Barnum1998,Barnum2000,Devetak2005}:
\begin{align}
C_{\textrm{Q}}(\mathcal{N}) = Q_{\textrm{reg}}(\mathcal{N}) \equiv \lim_{N\rightarrow \infty}  \frac{1}{N} \max_{\hat{\rho}} I_{\textrm{c}}( \mathcal{N}^{\otimes N}, \hat{\rho} ).
\end{align}
In the energy-constrained case, the coherent information should be optimized over all input states that satisfy an energy-constraint, such that at most $\bar{n}$ mean photons are fed to the channel in each use.

The bosonic pure-loss channel $\mathcal{N}[\eta,0]$ with a transmissivity $\eta\in [0,1]$ (or loss probability $\gamma\equiv 1-\eta$) is either degradable ($\eta\in(1/2,1]$) or anti-degradable ($\eta\in [0,1/2]$)\cite{Caruso2006}. Therefore, the regularization of its coherent information is unnecessary\cite{Devetak2005C,Caruso2006,Yard2008} and the optimal input state subject to an average photon number constraint $\textrm{Tr}[\hat{n}_{k} \hat{\rho} ]\le \bar{n}$ for all $k\in \lbrace 1,\cdots, N \rbrace$ is shown to be the single-mode thermal state $\hat{\tau}(\bar{n})$\cite{Wolf2007,Wilde2012} (see also Ref. \citen{Noh2019}):
\begin{align}
C_{\textrm{Q}}^{\le \bar{n}}(\mathcal{N}[\eta,0]) &= \lim_{N\rightarrow \infty}  \frac{1}{N} \max_{\hat{\rho}:  \textrm{Tr}[\hat{n}_{k} \hat{\rho} ]\le \bar{n} \forall k }  I_{\textrm{c}}( \mathcal{N}[\eta,0]^{\otimes N}, \hat{\rho} )
\nonumber\\
&= I_{\textrm{c}}( \mathcal{N}[\eta,0], \hat{\tau}(\bar{n}) ) = g(\eta \bar{n}) - ((1-\eta)\bar{n}),
\end{align}
where $g(x) \equiv S(\hat{\tau}(x)) =  (x+1)\log_{2}(x+1)-x\log_{2}x$ is the entropy of the thermal state $\hat{\tau}(x)$\cite{Bombelli1986}.

On the other hand, a general thermal-loss channel $\mathcal{N}[\eta,\nth]$ with a non-zero environmental thermal photon number $\nth \neq 0$ is neither degradable nor anti-degradable\cite{Caruso2006B,Holevo2007}. In this case, the single-mode thermal state $\hat{\tau}(\bar{n})$ is not necessarily the optimal input state and the associated coherent information (evaluated in Ref. \citen{Holevo2001}) only lower bounds the quantum capacity, i.e.,
\begin{align}
C_{\textrm{Q}}^{\le \bar{n}}(\mathcal{N}[\eta,\nth]) &\ge I_{\textrm{c}}( \mathcal{N}[\eta,\nth], \hat{\tau}(\bar{n}) )
\nonumber\\
&= g(\eta\bar{n}+(1-\eta)\nth)
\nonumber\\
&\quad -g\Big{(} \frac{D+(1-\eta)(\bar{n}-\nth)-1}{2} \Big{)}
\nonumber\\
&\quad -g\Big{(} \frac{D-(1-\eta)(\bar{n}-\nth)-1}{2} \Big{)}, \label{eq:thermal loss HW lower bound}
\end{align}
where $D\equiv
\sqrt{((1+\eta)\bar{n}+(1-\eta)\nth+1)^{2}-4\eta\bar{n}(\bar{n}+1)}$.
This is the best known lower bound for
$C_{\textrm{Q}}(\mathcal{N}[\eta,\nth])$ to date before our work. Below, we
demonstrate that correlated multi-mode thermal states can
outperform the single-mode thermal state of the same energy in the
noisy channel (near-zero capacity) regime. By doing so, we show
that higher quantum state transmission rates can be attained for
thermal-loss channels than previously believed.

\begin{theorem}
Consider a correlated $N$-mode thermal state $\hat{\mathcal{T}}(\boldsymbol{\mathrm{N}},\boldsymbol{\mathrm{n}})$ with $\boldsymbol{\mathrm{N}} = (M,N-M)$ and $\boldsymbol{\mathrm{n}} = (\frac{N}{M}\bar{n},0)$ and let $x = \frac{M}{N}$, where $M\in \lbrace 1,\cdots, N \rbrace$. Then, the coherent information with respect to the input state $\hat{\mathcal{T}}(\boldsymbol{\mathrm{N}},\boldsymbol{\mathrm{n}})$ is given by
\begin{align}
\frac{1}{N}I_{\textrm{c}}\big{(} \mathcal{N}[\eta,\nth]^{\otimes N} , \hat{\mathcal{T}}(\boldsymbol{\mathrm{N}},\boldsymbol{\mathrm{n}}) \big{)} = x I_{\textrm{c}}(\mathcal{N}[\eta,\nth], \hat{\tau}\Big{(}\frac{\bar{n}}{x}\Big{)} ). \label{eq:thermal loss correlated multi mode thermal state coherent information}
\end{align}
Since $x$ can be any rational number in $(0,1]$ and the set of rational numbers is a dense subset of the set of real numbers, we have the following improved lower bound of the quantum capacity of thermal-loss channels.
\begin{align}
C_{\textrm{Q}}^{\le \bar{n}}(\mathcal{N}[\eta,\nth]) &\ge \max_{0< x\le 1}x I_{\textrm{c}}(\mathcal{N}[\eta,\nth], \hat{\tau}\Big{(}\frac{\bar{n}}{x}\Big{)} ) .  \label{eq:thermal loss KN lower bound}
\end{align}
\label{theorem:Quantum capacity thermal loss}
\end{theorem}

The proof of Theorem \ref{theorem:Quantum capacity thermal loss} is given in Methods. Note that for $\boldsymbol{\mathrm{N}} = (M,N-M)$ and $\boldsymbol{\mathrm{n}} = (\frac{N}{M}\bar{n},0)$, $\textrm{Tr}[\hat{n}_{k} \hat{\mathcal{T}}(\boldsymbol{\mathrm{N}},\boldsymbol{\mathrm{n}}) ] = \bar{n}$ holds for all $k\in\lbrace1,\cdots, N\rbrace$ and thus $\hat{\mathcal{T}}(\boldsymbol{\mathrm{N}},\boldsymbol{\mathrm{n}})$ is a valid input state that satisfies the energy constraint. Also, our new bound in Eq. \eqref{eq:thermal loss KN lower bound} is at least as tight as the previous bound in Eq. \eqref{eq:thermal loss HW lower bound} since the previous bound can be recovered by plugging in $x=1$ to the objective function. Below, we show that in the noisy channel (near-zero capacity) regime, the optimal value of $x$ can be strictly less than $1$ and thus our bound is strictly tighter than the previous bound. We also explain this behavior in an intuitive manner by using simple mathematical concepts such as convexity of a function and convex hull of a non-convex region.

\begin{figure}
\centering
\includegraphics[width=18.0cm]{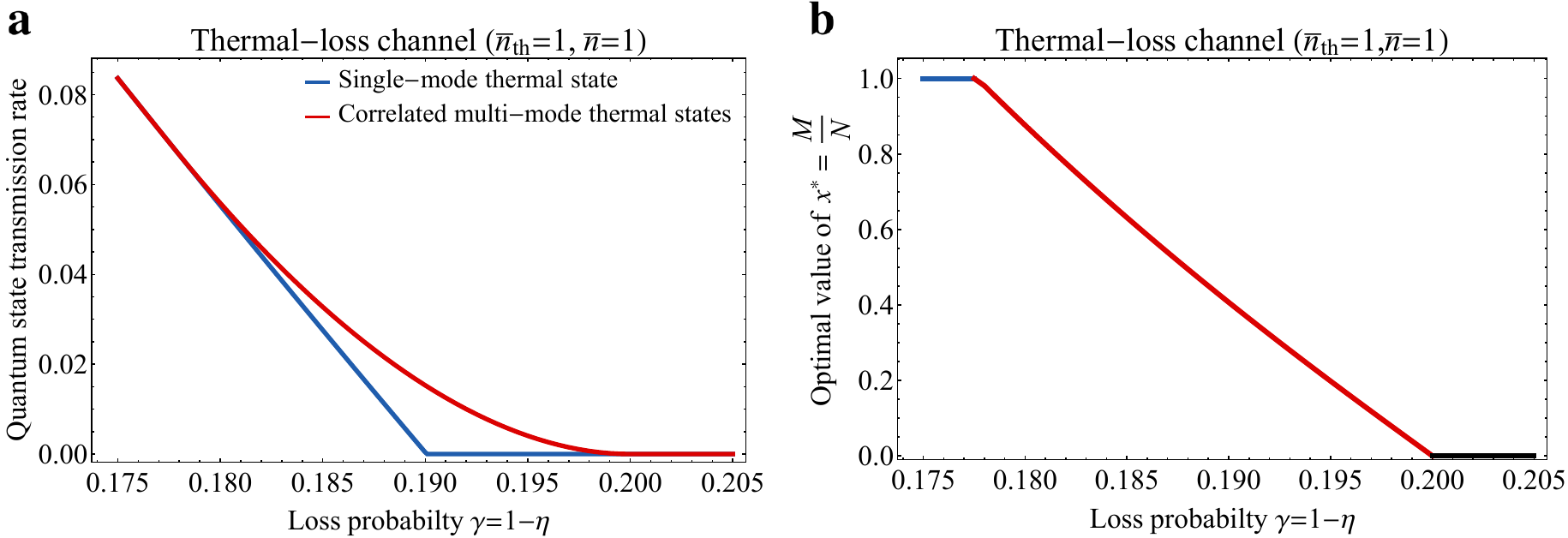}
\caption{\textbf{Achievable quantum state transmission rate of thermal-loss channels.} (\textbf{a}) Quantum state transmission rate of thermal-loss channels $\mathcal{N}[\eta,\nth=1]$ as a function of the loss probability $\gamma=1-\eta$ achievable with the single-mode thermal state $\hat{\tau}(\bar{n})$ (blue, Eq. \eqref{eq:thermal loss HW lower bound}) and with a correlated multi-mode thermal state $\hat{\mathcal{T}}(\boldsymbol{\mathrm{N}},\boldsymbol{\mathrm{n}})$ (red, Eq. \eqref{eq:thermal loss KN lower bound}) subject to the maximum allowed average photon number $\bar{n}=1$ per channel use. For the correlated multi-mode thermal states, the achievable rate was evaluated by taking $M/N = x^{\star}= \textrm{argmax}_{0 < x\le 1} x I_{\textrm{c}}(\mathcal{N}[\eta,\nth], \hat{\tau}(\bar{n}/x) )$, where $\boldsymbol{\mathrm{N}} = (M,N-M)$ and $\boldsymbol{\mathrm{n}}=(\frac{N}{M}\bar{n},0)$. (\textbf{b}) The optimal value of $x^{\star} = M/N$ at each $\gamma$, color-coded by the type of optimizer (blue: single-mode thermal states; red: correlated multi-mode thermal states), that yields the maximum quantum state transmission rate. We set $x^{\star}=M/N=0$ when all the states we consider yield vanishing quantum state transmission rate (black).  
}
\label{fig:Quantum capacity thermal loss}
\end{figure}

To demonstrate that our new bound can be strictly tighter than the previous bound, we take a family of thermal-loss channels $\mathcal{N}[\eta,\nth]$ with $\nth=1$ and compute the new bound in Eq. \eqref{eq:thermal loss KN lower bound} for each $\eta=1-\gamma$, assuming that the maximum allowed average photon number per channel is $\bar{n}=1$. In Fig. \ref{fig:Quantum capacity thermal loss}a, we plot the quantum state transmission rates achievable with the single-mode thermal state $\hat{\tau}(\bar{n}=1)$ and with the correlated multi-mode thermal states $\hat{\mathcal{T}}(\boldsymbol{\mathrm{N}},\boldsymbol{\mathrm{n}})$. When the loss probability is low (i.e., $\gamma \le 0.1775$), the single-mode thermal state yields the largest coherent information. However, when the loss probability is higher ($\gamma\ge 0.1775$), there exists a correlated multi-mode thermal state that outperforms the single-mode thermal state. Thus, we established a tighter lower bound to the quantum capacity of thermal-loss channels than previously known\cite{Holevo2001}. In Fig. \ref{fig:Quantum capacity thermal loss}b, we plot the optimal value of $M/N$ as a function of $\gamma$ that allows such a higher communication rate. It is important to note that only a finite number of modes is required if the optimal value of $x$ is a rational number. For example, $x^{\star}=3/8$ corresponds to the correlated $8$-mode thermal state $\hat{\mathcal{T}}(\boldsymbol{\mathrm{N}},\boldsymbol{\mathrm{n}})$ with $\boldsymbol{\mathrm{N}}=(M,N-M)=(3,5)$ and $\boldsymbol{\mathrm{n}} = (8\bar{n}/3,0)$. On the other hand, if $x^{\star}$ is irrational, one needs infinitely many modes to accurately obtain the rate $x I_{\textrm{c}}(\mathcal{N}[\eta,\nth], \hat{\tau} ( \bar{n}/x )  )|_{x=x^{\star}}$.

\subsection{Convexity of coherent information and superadditivity}

\begin{figure}
\centering
\includegraphics[width=18.0cm]{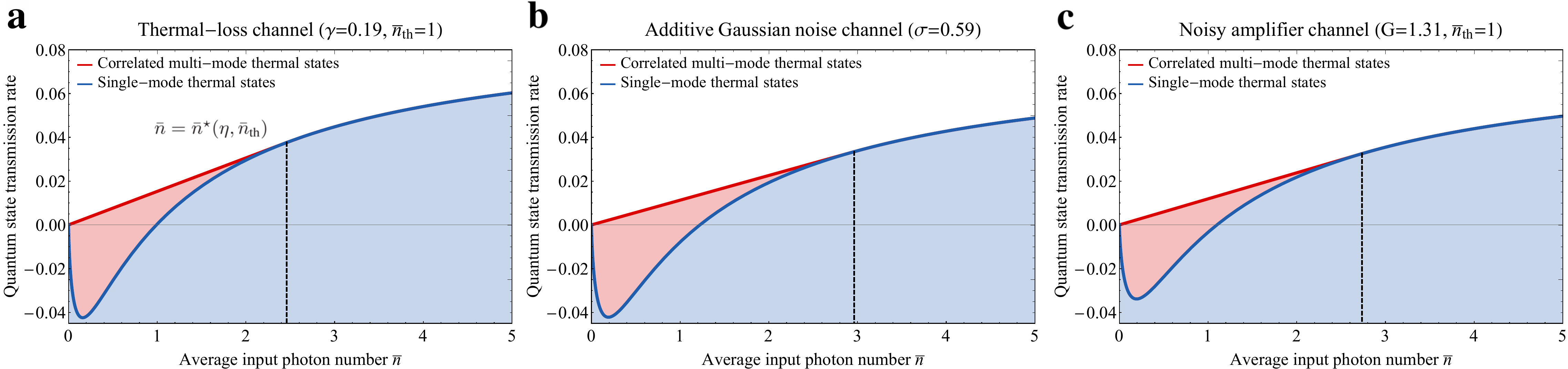}
\caption{\textbf{Convexity of coherent information and superadditivity.} Achievable quantum state transmission rate of the single-mode (blue) and correlated multi-mode (red) thermal states as a function of $\bar{n}$ for (\textbf{a}) the thermal-loss channel $\mathcal{N}[\eta=0.81,\nth = 1]$ (\textbf{b}) the additive Gaussian noise channel $\mathcal{N}_{\textrm{B}_{2}}[\sigma  = 0.59]$ and (\textbf{c}) the noisy amplifier channel $\mathcal{A}[G=1.31,\nth=1]$. Note that in (\textbf{a}), the first-order contact point is given by $\bar{n}^{\star}(\eta=0.81,\nth=1) = 2.458$ which corresponds to $x^{\star} = 1/2.458 = 0.407$. This value agrees with $x^{\star}=0.407$ which is independently obtained in Fig. \ref{fig:Quantum capacity thermal loss}b for $\gamma=0.19$ and $\nth = 1$ (see also the main text).  
}
\label{fig:achievable rates verses nbar}
\end{figure}

We now explain the non-trivial behavior shown in Fig. \ref{fig:Quantum capacity thermal loss} (i.e., $x^{\star} < 1$) in an intuitive way. Specifically, we relate the observed non-trivial behavior with the convexity of the coherent information $I_{\textrm{c}}(\mathcal{N}[\eta,\nth],\hat{\tau}(\bar{n}))$ in the allowed average photon number $\bar{n}$ for fixed values of $\eta$ and $\nth$. For concreteness, we take the thermal-loss channel $\mathcal{N}[\eta,\nth]$ with $\eta = 0.81$ (or $\gamma = 0.19$) and $\nth = 1$ and plot its coherent information $I_{\textrm{c}}(\mathcal{N}[\eta,\nth],\hat{\tau}(\bar{n}))$ with respect to single-mode thermal states $\hat{\tau}(\bar{n})$ as a function of $\bar{n}$. As can be seen from the solid blue line in Fig. \ref{fig:achievable rates verses nbar}a, the coherent information $I_{\textrm{c}}(\mathcal{N}[\eta,\nth],\hat{\tau}(\bar{n}))$ is convex in $\bar{n}$ for small $\bar{n}$ and concave for large $\bar{n}$. Consider the region of rates achievable by the single-mode thermal states $A^{(1)}_{\eta,\nth} \equiv \lbrace (\bar{n},R) | \bar{n}\ge 0  \textrm{ and } R \le I_{\textrm{c}}(\mathcal{N}[\eta,\nth],\hat{\tau}(\bar{n}))  \rbrace$ (shaded blue region in Fig. \ref{fig:achievable rates verses nbar}a) and also its convex hull $A^{(\infty)}_{\eta,\nth} \equiv \textrm{ConvexHull}(A^{(1)}_{\eta,\nth})$ (shaded red and blue regions in Fig. \ref{fig:achievable rates verses nbar}a). We observe that the region $A^{(\infty)}_{\eta,\nth}$ is achievable by correlated multi-mode thermal states: Consider a generic convex combination of $r$ points in $A^{(1)}_{\eta,\nth}$, i.e.,
\begin{align}
\sum_{k=1}^{r} \lambda_{k} \Big{(} \bar{n}_{k} , I_{\textrm{c}}( \mathcal{N}[\eta,\nth],\hat{\tau}(\bar{n}_{k}) )  \Big{)},
\end{align}
where $\lambda_{k}\ge 0$ for all $k\in\lbrace 1,\cdots, r\rbrace$ and $\sum_{k=1}^{r} \lambda_{k}=1$. Then, the rate $\sum_{k=1}^{r} \lambda_{k} I_{\textrm{c}}( \mathcal{N}[\eta,\nth],\hat{\tau}(\bar{n}_{k}) )$
can be achieved by a correlated multi-mode thermal state $\hat{\mathcal{T}}(\boldsymbol{\mathrm{N}},\boldsymbol{\mathrm{n}})$ with $\boldsymbol{\mathrm{N}} = (N_{1},\cdots, N_{r})$ and $\boldsymbol{\mathrm{n}} = (\bar{n}_{1},\cdots, \bar{n}_{r})$ such that $\lambda_{k} = N_{k}/N$ for all $k\in \lbrace1,\cdots, r\rbrace$ where $N=\sum_{k=1}^{r}N_{k}$. Note that $\lambda_{k}$ should be a rational number. Similarly as above, however, by choosing a sufficiently large $N$ one can approximate any irrational $\lambda_{k}$ to a desired accuracy which can be arbitrarily small.    



Importantly, due to the convexity of the coherent information $I_{\textrm{c}}(\mathcal{N}[\eta,\nth],\hat{\tau}(\bar{n}))$ in the small $\bar{n}$ regime, the region $A^{(\infty)}_{\eta,\nth}$ properly contains the region $A^{(1)}_{\eta,\nth}$, as indicated by the shaded red region in Fig. \ref{fig:achievable rates verses nbar}a. This is why correlated multi-mode thermal states outperform single-mode thermal states in the noisy channel regime. In particular, the highest achievable rate can be obtained by taking the convex combination of the origin $(0,0)$ and the first-order contact point $(\bar{n}^{\star}(\eta,\nth) ,  I_{\textrm{c}}(\mathcal{N}[\eta,\nth], \hat{\tau} ( \bar{n}^{\star}(\eta,\nth) ) )$ with some weights $\lambda$ and $1-\lambda$, respectively (see the solid red line in Fig. \ref{fig:achievable rates verses nbar}). Note that the rate $x I_{\textrm{c}}(\mathcal{N}[\eta,\nth], \hat{\tau} ( \bar{n}/x )  )$ in Eq. \eqref{eq:thermal loss KN lower bound} can be understood as the one that is derived from such a convex combination with $1-\lambda = x = \bar{n}/\bar{n}^{\star}(\eta,\nth)$. For example, in the case of thermal-loss channel $\mathcal{N}[\eta,\nth]$ with $\eta = 0.81$ (or $\gamma = 0.19$) and $\nth =1$, the first-order contact point is given by $\bar{n}^{\star}(\eta,\nth) = 2.458$ (see Fig. \ref{fig:achievable rates verses nbar}) which corresponds to $x = 0.407$ for $\bar{n}=1$: This agrees with the optimal value $x^{\star} = 0.407$ in Fig. \ref{fig:Quantum capacity thermal loss}b for $\eta = 0.81$ (or $\gamma = 0.19$), $\nth=1$, and $\bar{n}=1$.

We remark that the coherent information (with respect to single-mode thermal states) of other Gaussian channels that are neither degradable nor anti-degradable, such as additive Gaussian noise channels and noisy amplification channels, also exhibit a non-trivial behavior similarly to the thermal-loss channels (see Methods for the definition of these other channels). More specifically, the coherent information of the additive Gaussian noise channel $\mathcal{N}_{\textrm{B}_{2}}[\sigma]$ is given by
\begin{align}
I_{\textrm{c}}(\mathcal{N}_{\textrm{B}_{2}}[\sigma] , \hat{\tau}(\bar{n}) ) &= g(\bar{n}+\sigma^{2}) - g\Big{(} \frac{D' + \sigma^{2} -1}{2} \Big{)}
\nonumber\\
&\quad -g\Big{(} \frac{D' - \sigma^{2} -1}{2} \Big{)} ,
\end{align}
where $D' \equiv \sqrt{ (2\bar{n}+\sigma^{2} + 1)^{2} - 4\bar{n}(\bar{n}+1) }$. Also, the coherent information of the noisy amplifier channel $\mathcal{A}[G,\nth]$ is given by
\begin{align}
I_{\textrm{c}}(\mathcal{A}[G,\nth] , \hat{\tau}(\bar{n}) ) &= g(G\bar{n}+(G-1)(\nth+1))
\nonumber\\
&\quad -g\Big{(} \frac{D'' + (G-1)(\bar{n}+\nth+1) -1 }{2} \Big{)}
\nonumber\\
&\quad -g\Big{(} \frac{D'' - (G-1)(\bar{n}+\nth+1) -1 }{2} \Big{)},
\end{align}
where \begin{equation}D''\equiv \sqrt{ ( (G+1)\bar{n} +
(G-1)(\nth+1) + 1 )^{2} - 4G\bar{n}(\bar{n}+1) }.\end{equation} As
can be seen from Fig. \ref{fig:achievable rates verses nbar}b and
Fig. \ref{fig:achievable rates verses nbar}c, the coherent
information of these other channels also exhibit the same convex
behavior in the small $\bar{n}$ regime. Therefore, higher quantum
state transmission rates can be achieved for these other Gaussian
channels as well by using the correlated multi-mode thermal states
instead of using the single-mode thermal states, analogously to
the case of thermal-loss channels as shown here.

\subsection{Reverse coherent information and two-way quantum capacity}

We show that a similar technique can be used to establish an improved lower bound of the two-way quantum capacity of thermal-loss channels. Let $I_{\textrm{rc}}(\mathcal{N},\hat{\rho})$ be the reverse coherent information of a bosonic channel $\mathcal{N}$ with respect to an input state $\hat{\rho}$\cite{Pirandola2009}.
\begin{align}
I_{\textrm{rc}}(\mathcal{N},\hat{\rho}) \equiv S(\hat{\rho}) - S\big{(} \mathcal{N}^{\textrm{c}}(\hat{\rho}) \big{)}.
\end{align}
Both the coherent information and the reverse coherent information of a channel $\mathcal{N}$ are lower bounds of the channel's two-way quantum capacity:
\begin{align}
C_{\textrm{Q},\leftrightarrow}(\mathcal{N}) \ge \max_{\hat{\rho}} \Big{[} \max\big{(} I_{\textrm{c}}(\mathcal{N},\hat{\rho}) , I_{\textrm{rc}}(\mathcal{N},\hat{\rho}) \big{)} \Big{]} .
\end{align}
In the energy constrained cases, the maximization should be performed over all input states that satisfy the energy constraint.

The best known lower bound (before our work) of the two-way
quantum capacity of a thermal-loss channel is either the channel's
coherent information or reverse coherent information with respect
to a single-mode thermal state, i.e.,
\begin{align}
C^{\le \bar{n}}_{\textrm{Q},\leftrightarrow}(\mathcal{N}[\eta,\nth]) \ge \begin{cases}
 I_{\textrm{c}} ( \mathcal{N}[\eta,\nth],\hat{\tau}(\bar{n}) ) & \bar{n}\le \nth \\
  I_{\textrm{rc}} ( \mathcal{N}[\eta,\nth],\hat{\tau}(\bar{n}) ) & \bar{n}> \nth
\end{cases}, \label{eq:thermal loss two way folklore lower bound}
\end{align}
where $I_{\textrm{c}} ( \mathcal{N}[\eta,\nth],\hat{\tau}(\bar{n}) )$ is
given in Eq. \eqref{eq:thermal loss HW lower bound} and $I_{\textrm{rc}} (
\mathcal{N}[\eta,\nth],\hat{\tau}(\bar{n}) )$ can be obtained by
replacing $g(\eta \bar{n} + (1-\eta)\nth)$ in Eq.
\eqref{eq:thermal loss HW lower bound} by
$g(\bar{n})$\cite{Pirandola2009,Pirandola2017}, i.e.,
\begin{align}
I_{\textrm{c}} ( \mathcal{N}[\eta,\nth],\hat{\tau}(\bar{n}) ) &= g(\bar{n})  -g\Big{(} \frac{D+(1-\eta)(\bar{n}-\nth)-1}{2} \Big{)}
\nonumber\\
&\quad -g\Big{(} \frac{D-(1-\eta)(\bar{n}-\nth)-1}{2} \Big{)}.
\end{align}
In the special case where $\nth=0$ and $\bar{n}\rightarrow
\infty$, the lower bound in Eq. \eqref{eq:thermal loss two way
folklore lower bound} is given by
$-\log_{2}(1-\eta)$\cite{Pirandola2009} and coincides with the
upper bound established in Ref. \citen{Pirandola2017}. Except for
this special case, it is an open question whether the lower bound
in Eq. \eqref{eq:thermal loss two way folklore lower bound} equals
the true two-way quantum capacity of thermal-loss channels: Here,
we provide a negative answer to this question by showing that
higher two-way quantum state transmission rate can be achieved by
using correlated multi-mode thermal states.

\begin{theorem}
Consider a correlated $N$-mode thermal state
$\hat{\mathcal{T}}(\boldsymbol{\mathrm{N}},\boldsymbol{\mathrm{n}})$ with $\boldsymbol{\mathrm{N}} = (M,N-M)$ and
$\boldsymbol{\mathrm{n}} = (\bar{n}_{1},\bar{n}_{2})$ such that $M\bar{n}_{1}+(N-M)\bar{n}_{2} = N\bar{n}$ and
let $x= \frac{M}{N}$, where $M\in\lbrace 1,\cdots, N \rbrace$. The
following two-way quantum state transmission rate can be achieved
by using $\hat{\mathcal{T}}(\boldsymbol{\mathrm{N}},\boldsymbol{\mathrm{n}})$ and hybridizing
forward and backward strategies:
\begin{align}
\frac{M}{N}I_{\textrm{c}}(\mathcal{N}[\eta,\nth], \hat{\tau}(\bar{n}_{1})) + \frac{N-M}{N}I_{\textrm{rc}}(\mathcal{N}[\eta,\nth], \hat{\tau}(\bar{n}_{2})). \label{eq:thermal loss correlated multi mode thermal state two-way rate}
\end{align}
Since $x$ can be any rational number in $(0,1]$ and the set of rational numbers is a dense subset of the set of real numbers, we have the following improved lower bound for the two-way energy-constrained quantum capacity of a thermal-loss channel:
\begin{align}
C_{\textrm{Q},\leftrightarrow}^{\le \bar{n}}(\mathcal{N}[\eta,\nth]) &\ge \max_{x,\bar{n}_{1},\bar{n}_{2}} \Big{[} x I_{\textrm{c}}(\mathcal{N}[\eta,\nth], \hat{\tau}(\bar{n}_{1}) )
\nonumber\\
&\qquad + (1-x)I_{\textrm{rc}}(\mathcal{N}[\eta,\nth], \hat{\tau}(\bar{n}_{2}) )  \Big{]}  , \label{eq:thermal loss two way KN lower bound}
\end{align}
where the maximization is performed subject to $0\le x\le 1$ and $x\bar{n}_{1} + (1-x)\bar{n}_{2} =\bar{n}$.  \label{theorem:two-way quantum capacity thermal loss}
\end{theorem}
The proof of Theorem~\ref{theorem:two-way quantum capacity thermal
loss} is given in Methods. Our new bound in Eq. \eqref{eq:thermal
loss two way KN lower bound} is at least as tight as the previous
bound in Eq. \eqref{eq:thermal loss two way folklore lower bound}
because the previous bound can be realized by plugging in $x=1$
and $\bar{n}_{1} = \bar{n}$ or $x=0$ and $\bar{n}_{2} = \bar{n}$. Moreover, we show that in the noisy channel
(near-zero capacity) regime, correlated multi-mode thermal states
outperform single-mode thermal states of the same energy and thus
our bound is strictly tighter than the previous bound.

\begin{figure}
\centering
\includegraphics[width=18.0cm]{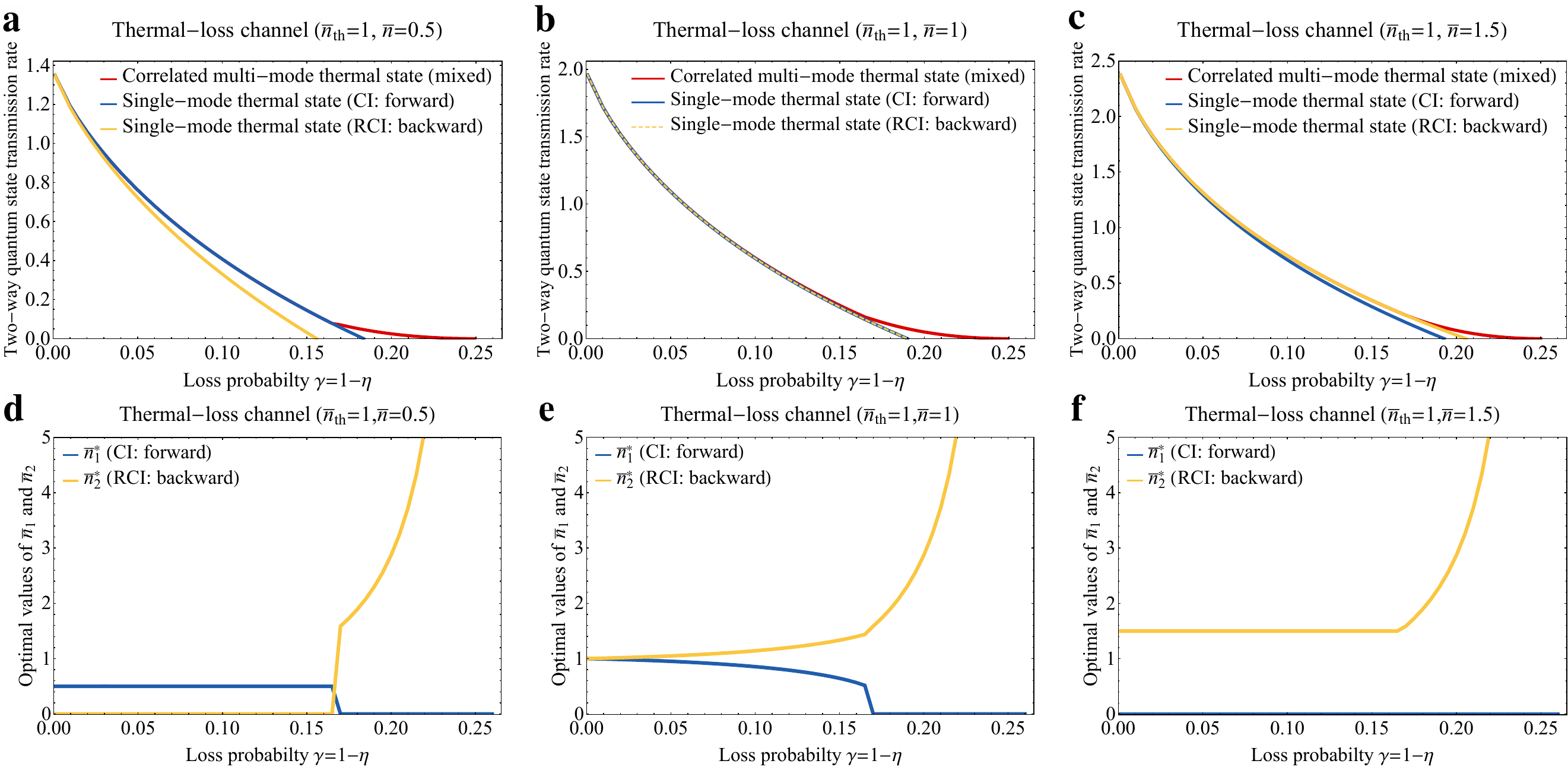}
\caption{\textbf{Achievable two-way quantum state transmission rate of thermal loss channels.} We plot the achievable two-way quantum state transmission rate of a thermal-loss channel $\mathcal{N}[\eta,\nth=1]$ subject to the maximum allowed average photon number (\textbf{a}) $\bar{n}=0.5$ (\textbf{b}) $\bar{n}=1$ (\textbf{c}) $\bar{n}=1.5$ per channel use. In (\textbf{a})–(\textbf{c}), the blue and yellow lines respectively represent the coherent information $I_{\textrm{c}}(\mathcal{N}[\eta,\nth],\hat{\tau}(\bar{n}))$ and the reverse coherent information $I_{\textrm{rc}}(\mathcal{N}[\eta,\nth],\hat{\tau}(\bar{n}))$ with respect to single-mode thermal states. The achievable two-way quantum state transmission rate of correlated multi-mode thermal states (red lines in (\textbf{a})–(\textbf{c})) was evaluated by taking $(x^{\star},\bar{n}_{1}^{\star},\bar{n}_{2}^{\star})  = \textrm{argmax}_{x,\bar{n}_{1},\bar{n}_{2}} [ x I_{\textrm{c}}(\mathcal{N}[\eta,\nth], \hat{\tau}(\bar{n}_{1}) )  + (1-x) I_{\textrm{rc}}(\mathcal{N}[\eta,\nth], \hat{\tau}(\bar{n}_{2}) )    ]$ subject to $0 \le x \le 1$ and $x\bar{n}_{1}+(1-x)\bar{n}_{2} = \bar{n}$ (see Theorem \ref{theorem:two-way quantum capacity thermal loss}). In (\textbf{d})–(\textbf{f}), the optimal values $\bar{n}_{1}^{\star}$ and $\bar{n}_{2}^{\star}$ are respectively represented by the blue and yellow lines for (\textbf{d}) $\bar{n}=0.5$ (\textbf{e}) $\bar{n}=1$ (\textbf{f}) $\bar{n}=1.5$. The optimal value $x^{\star}$ can be obtained by evaluating $x^{\star} = (\bar{n}-\bar{n}_{2}^{\star})/(\bar{n}_{1}^{\star}-\bar{n}_{2}^{\star})$. 
}
\label{fig:two-way quantum capacity thermal loss}
\end{figure}

To demonstrate that our new bound can be strictly tighter than the previous bound, we take a family of thermal-loss channels $\mathcal{N}[\eta,\nth]$ with $\nth =1$. Then, we compute the new bound in Eq. \eqref{eq:thermal loss two way KN lower bound} for each $\eta= 1-\gamma$ for three different maximum allowed average photon numbers per channel use, i.e., $\bar{n}=0.5$ (Fig. \ref{fig:two-way quantum capacity thermal loss}a), $\bar{n}=1$ (Fig. \ref{fig:two-way quantum capacity thermal loss}b), and $\bar{n}=1.5$ (Fig. \ref{fig:two-way quantum capacity thermal loss}c). As can be seen from the top panel of Fig. \ref{fig:two-way quantum capacity thermal loss}, the coherent information $I_{\textrm{rc}}(\mathcal{N}[\eta,\nth],\hat{\tau}(\bar{n}))$ (blue lines) is larger than, equal to, and smaller than the reverse coherent information $I_{\textrm{rc}}(\mathcal{N}[\eta,\nth],\hat{\tau}(\bar{n}))$ (yellow lines) for $\bar{n} = 0.5$, $\bar{n} = 1$, and $\bar{n}=1.5$, respectively. In all cases, our new bound obtained by using correlated multi-mode thermal states (red lines) can be strictly tighter than the previous bound in the large loss probability regime where the two-way quantum capacity almost vanishes. In this regime, the best two-way quantum state transmission rate is achieved by mixing forward (coherent information) and backward (reverse coherent information) strategies, as can be seen from the bottom panels of Fig. \ref{fig:two-way quantum capacity thermal loss}.

\subsection{Private information and private capacity}

Lastly, we apply our general technique to improve the lower bound of the energy-constrained private capacity of the thermal-loss channel. Consider a classical-quantum state $\hat{\sigma} = \int dx p(x)|x\rangle\langle x|\otimes \hat{\rho}_{x}$ where $\langle x|x'\rangle = \delta(x-x')$ and $\int dxp(x)=1$ and let $\hat{\rho} \equiv \int dx p(x) \hat{\rho}_{x}$. The private capacity of a quantum channel characterizes the channel's maximum achievable secure classical communication rate. The private information of channel $\mathcal{N}$ with respect to the classical-quantum state $\hat{\sigma}$ is defined as
\begin{align}
I_{\textrm{p}}(\mathcal{N}, \hat{\sigma}) &\equiv S\big{(} \mathcal{N}(\hat{\rho}) \big{)} - S\big{(} \mathcal{N}^{\textrm{c}}(\hat{\rho}) \big{)}
\nonumber\\
&\quad - \int dx p(x) \Big{[} S\big{(} \mathcal{N}( \hat{\rho}_{x} ) \big{)} - S\big{(} \mathcal{N}^{\textrm{c}}( \hat{\rho}_{x} ) \big{)} \Big{]} .
\end{align}
The private capacity $C_{\textrm{P}}(\mathcal{N})$ of a quantum channel
$\mathcal{N}$ is equal to the channel's regularized private
information $P_{\textrm{reg}}(\mathcal{N})$\cite{Devetak2005}:
\begin{align}
C_{\textrm{P}}(\mathcal{N}) = P_{\textrm{reg}}(\mathcal{N}) \equiv \lim_{N\rightarrow \infty} \frac{1}{N} \max_{\hat{\sigma}} I_{\textrm{p}}(\mathcal{N}^{\otimes N} , \hat{\sigma} ).
\end{align}
In the energy-constrained case, the maximization should be
performed over all classical-quantum states $\hat{\sigma}=\int dx
p(x)|x\rangle\langle x|\otimes \hat{\rho}_{x}$ such that
$\hat{\rho} = \int dx p(x) \hat{\rho}_{x}$ satisfies the energy
constraint.

The quantum capacity of a channel $\mathcal{N}$ is always a lower
bound of the channel's private capacity\cite{Devetak2005} and thus
the coherent information
$I_{\textrm{c}}(\mathcal{N}[\eta,\nth],\hat{\tau}(\bar{n}))$ in Eq.
\eqref{eq:thermal loss HW lower bound} is also a lower bound of
the private capacity of the thermal-loss channel
$\mathcal{N}[\eta,\nth]$. Correspondingly, our new bound of the
quantum capacity in Theorem \ref{theorem:Quantum capacity thermal
loss} is also a valid lower bound of the private capacity which
can be strictly tighter than
$I_{\textrm{c}}(\mathcal{N}[\eta,\nth],\hat{\tau}(\bar{n}))$. However, it
was shown that higher secure classical communication rate (than
the coherent information
$I_{\textrm{c}}(\mathcal{N}[\eta,\nth],\hat{\tau}(\bar{n}))$) can be
achieved by using an ensemble of displaced thermal
states\cite{Sharma2018}. More specifically, by using a
classical-quantum state
\begin{align}
\hat{\sigma}(\bar{n}_{1},\bar{n}_{2}) \equiv \int d^{2}\alpha \frac{ e^{-|\alpha|^{2}/\bar{n}_{1}} }{\pi \bar{n}_{1}}  |\alpha\rangle\langle\alpha| \otimes \hat{D}(\alpha) \hat{\tau}(\bar{n}_{2}) \hat{D}^{\dagger}(\alpha) \label{eq:classical-quantum state KS}
\end{align}
such that $\bar{n}_{1}+\bar{n}_{2} = \bar{n}$, where $\alpha = \alpha_{\textrm{R}}+i\alpha_{\textrm{I}}$ and $\langle \alpha |\alpha'\rangle = \delta^{(2)}(\alpha-\alpha')$, the private communication rate
\begin{align}
I_{\textrm{p}}(\mathcal{N}[\eta,\nth],\hat{\sigma}) &= I_{\textrm{c}}(\mathcal{N}[\eta,\nth], \hat{\tau}(\bar{n}_{1} + \bar{n}_{2} ) )
\nonumber\\
&\quad - I_{\textrm{c}}(\mathcal{N}[\eta,\nth], \hat{\tau}(\bar{n}_{2} ) )
\end{align}
can be achieved. Thus, we have the following lower bound of the energy-constrained private capacity of thermal-loss channels:
\begin{align}
C_{\textrm{P}}^{\le \bar{n}}(\mathcal{N}[\eta,\nth]) &\ge f(\eta,\nth,\bar{n})
\nonumber\\
& \equiv  \max_{0\le \bar{n}_{2} \le \bar{n} }  \Big{[}  I_{\textrm{c}}(\mathcal{N}[\eta,\nth], \hat{\tau}( \bar{n}  ) )
\nonumber\\
&\qquad\qquad\quad - I_{\textrm{c}}(\mathcal{N}[\eta,\nth], \hat{\tau}(\bar{n}_{2} ) ) \Big{]}. \label{eq:thermal loss private KS lower bound}
\end{align}
Since the coherent information $I_{\textrm{c}}(\mathcal{N}[\eta,\nth],
\hat{\tau}( \bar{n}  ) )$ can be recovered by plugging in
$\bar{n}_{2}=0$, this bound is at least as tight as the coherent
information bound. In the noisy channel regime, the bound in Eq.
\eqref{eq:thermal loss private KS lower bound} is strictly tighter
than the coherent information bound (see the blue and yellow lines
in Fig. \ref{fig:secure classical communication rate thermal
loss}a). Moreover, it is also strictly larger than our new bound
for the quantum capacity in Eq. \eqref{eq:thermal loss KN lower
bound} (see the red and yellow lines in Fig. \ref{fig:secure
classical communication rate thermal loss}a). Therefore, our new
bound for the quantum capacity is not the tightest lower bound for
the private capacity. Nevertheless, we show below that our
general technique can also be used to further improve the bound in
Eq.~\eqref{eq:thermal loss private KS lower bound}. We establish the following result. 

\begin{theorem}
The energy-constrained private capacity of a thermal-loss channel is lower bounded as follows:
\begin{align}
C_{\textrm{P}}^{\le \bar{n}}(\mathcal{N}[\eta,\nth]) &\ge F(\eta,\nth,\bar{n}) \equiv \max_{0<x\le 1} xf\Big{(} \eta,\nth,\frac{\bar{n}}{x} \Big{)}, \label{eq:thermal loss private KN lower bound}
\end{align}
where $f(\eta,\nth,\bar{n})$ is defined in Eq. \eqref{eq:thermal loss private KS lower bound}.
\label{theorem:private capacity thermal loss}
\end{theorem}

\begin{figure}
\centering
\includegraphics[width=18.0cm]{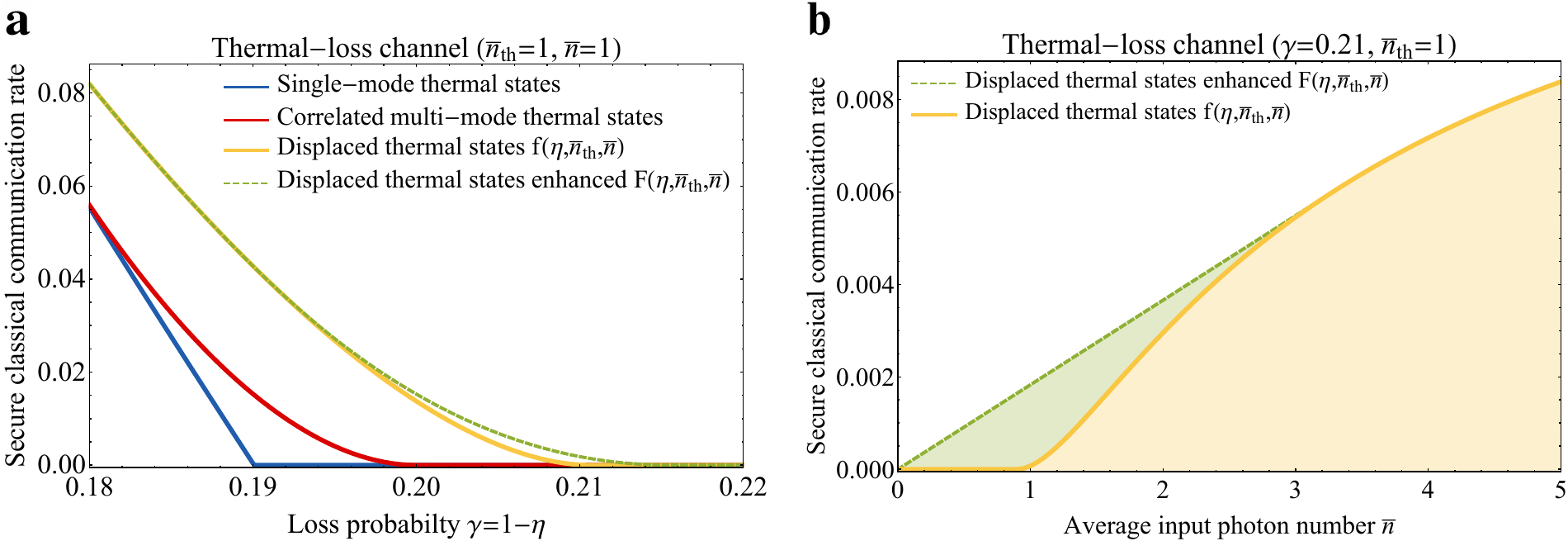}
\caption{\textbf{Achievable private communication rate of thermal-loss channels}. (\textbf{a}) Achievable private communication rate of a thermal-loss channel $\mathcal{N}[\eta,\nth]$ with $\nth=1$ subject to the maximum allowed average photon number $\bar{n}=1$ per channel use. The blue and red lines represent the achievable rates with the single-mode thermal state (Eq. \eqref{eq:thermal loss HW lower bound}) and correlated multi-mode thermal states (Eq. \eqref{eq:thermal loss KN lower bound}), respectively, and are identical to the blue and red lines in Fig. \ref{fig:Quantum capacity thermal loss}a. The yellow line represents the lower bound of Ref. \citen{Sharma2018} ($f(\eta,\nth,\bar{n})$; see also Eq. \eqref{eq:thermal loss private KS lower bound}) obtained by using the displaced thermal state in Eq. \eqref{eq:classical-quantum state KS}. The dashed green line represents our improved lower bound ($F(\eta,\nth,\bar{n})$; see also Eq. \eqref{eq:thermal loss private KN lower bound}) obtained by using the classical-quantum state we constructed in Eq. \eqref{eq:classical-quantum state KN}. (\textbf{b}) Achievable private communication rate of a thermal-loss channel $\mathcal{N}[\eta,\nth]$ with $\eta=0.79$ (or $\gamma=0.21$) and $\nth =1$ as a function of the maximum allowed average photon number $\bar{n}$. The yellow line represents the bound of Ref. \citen{Sharma2018} ($f(\eta,\nth,\bar{n})$) and the dashed green line represents our bound ($F(\eta,\nth,\bar{n})$). Similarly as in the case of quantum state transmission (Figs. \ref{fig:Quantum capacity thermal loss} and \ref{fig:achievable rates verses nbar}), the non-trivial advantage of our classical-quantum state in Eq. \eqref{eq:thermal loss private KN lower bound} is due to the fact that $f(\eta,\nth,\bar{n})$ is convex in the small $\bar{n}$ regime and thus the convex hull of the achievable region $\lbrace (\bar{n},R)| \bar{n}\ge 0 \textrm{ and }R\le f(\eta,\nth,\bar{n})  \rbrace$ properly contains the region itself (see also the main text).  
   }
\label{fig:secure classical communication rate thermal loss}
\end{figure}

The proof of Theorem \ref{theorem:private capacity thermal loss} is given in Methods. As can be seen from Fig. \ref{fig:secure classical communication rate thermal loss}a, our new bound in Eq. \eqref{eq:thermal loss private KN lower bound} is strictly tighter than the bound in Eq. \eqref{eq:thermal loss private KS lower bound} in the noisy channel regime. Similarly as above, this non-trivial advantage is due to the convexity of the function $f(\eta,\nth,\bar{n})$ in $\bar{n}$ in the small $\bar{n}$ regime for fixed values of $\eta$ and $\nth$. For illustration, we plot in Fig. \ref{fig:secure classical communication rate thermal loss}b the function $f(\eta,\nth,\bar{n})$ as a function of $\bar{n}$ for $\eta=0.79$ (or $\gamma=0.21$) and $\nth=1$. Due to the convexity of $f(\eta,\nth,\bar{n})$ in the small $\bar{n}$ regime, the convex hull of the achievable region $\lbrace (\bar{n},R)| \bar{n}\ge 0 \textrm{ and }R\le f(\eta,\nth,\bar{n})  \rbrace$ properly contains the region itself (see the shaded green region in Fig. \ref{fig:secure classical communication rate thermal loss}b) and this explains the superior performance of our new bound in Eq. \eqref{eq:thermal loss private KN lower bound}.


\section*{Discussion}

\noindent In this work, we have improved the lower bounds to the various energy-constrained quantum capacities of the thermal-loss channel, which is the basic model for realistic optical and microwave communication channels. In this way, our work shows that higher communication rates can be achieved for various quantum communication tasks with these practically relevant quantum channels than previously believed. Below, we make several remarks and discuss possible new research directions.

Firstly, it was shown in a related work\cite{Lupo2009} that a global encoding scheme with a correlated Gaussian input state can yield larger coherent information than a local encoding scheme with an uncorrelated Gaussian input state for lossy bosonic channels with correlated environmental noise. We remark that our work differs from this previous work in that we show a correlated Gaussian input state can outperform its uncorrelated counterpart even for the usual thermal-loss channels with uncorrelated environmental noise. Note that the loss model with uncorrelated environmental noise which we consider here has greater practical relevance because noise in realistic optical and microwave communication channels is well approximated by thermal-loss channels with uncorrelated environmental thermal noise\cite{Xiang2017,Axline2018}.

Additionally, our result in Theorem \ref{theorem:Quantum capacity thermal loss} can be understood as the establishment of the superadditivity of the coherent information of thermal-loss channels with respect to Gaussian input states: As shown in Ref. \citen{Holevo2001}, the single-mode thermal state $\hat{\tau}(\bar{n})$ is the optimal single-mode Gaussian input state for the coherent information of thermal-loss channels. Since we show that multi-mode correlated thermal states (which are Gaussian) sometimes outperform the single-mode thermal state, it means that the coherent information of thermal-loss channels is superadditive with respect to Gaussian input states. On the other hand, it is still unclear whether the coherent information of thermal-loss channels is genuinely superadditive with respect to all input states. This is because technically there is still a possibility that some non-Gaussian input state may outperform all Gaussian input states. We leave this optimality question in the non-Gaussian domain as an open research direction. 

Another interesting open question is whether the convexity
argument presented here can be adapted to explain the known
superadditivity behavior of the qubit
depolarization\cite{DiVincenzo1998,Smith2007,Fern2008} and
dephrasure\cite{Leditzky2018,Pirandola2019c,Bausch2018} channels.
To contrast, we remark that the coherent information of a
degradable channel is concave with respect to input states and its quantum capacity is
additive\cite{Devetak2005C,Caruso2006,Yard2008} (see also Ref. \citen{Watanabe2012}). 

We also remark that our improvement of the lower bounds is not strong enough to close the gap between the lower bounds and the best-known upper bounds of various energy-constrained quantum capacities of thermal-loss channels \cite{Pirandola2017,Sharma2018,Rosati2018,Noh2019}. It will thus be interesting to see whether it is possible to further improve the lower and upper bounds to get a better understanding of the various quantum capacities of thermal-loss channels.   

Finally, we emphasize that we did not provide explicit strategies to achieve various quantum communication rates established in this work but only proved their existence. This is because the achievability of the coherent information, reverse coherent information, and private information is based on random coding arguments. Therefore, it will be an interesting research avenue to look for explicit quantum communication protocols (for example by using GKP codes\cite{Gottesman2001,Harrington2001} or polar codes\cite{Arikan2009,Lacerda2017}) that can be implemented efficiently while achieving (or even improving) the rates we have established here.


\begin{methods}

\subsection{Gaussian states and channels}  A bosonic mode is described by its annihilation and creation operators $\hat{a}$ and $\hat{a}^{\dagger}$ that satisfy the commutation relation $[\hat{a},\hat{a}^{\dagger}] = 1$ (see, for example, Ref. \citen{Braunstein2005}). Let $\boldsymbol{\hat{\mathrm{x}}} \equiv (\hat{q}_{1},\cdots,\hat{q}_{N},\hat{p}_{1},\cdots,\hat{p}_{N})$ be the quadrature operators of $N$ bosonic modes where $\hat{q}_{k}\equiv (\hat{a}_{k}^{\dagger}+\hat{a}_{k})/\sqrt{2}$ and $\hat{p}_{k} \equiv i(\hat{a}_{k}^{\dagger}-\hat{a}_{k})/\sqrt{2}$. The quadrature operators satisfy the commutation relation $[\boldsymbol{\hat{\mathrm{x}}}_{j} , \boldsymbol{\hat{\mathrm{x}}}_{k} ] = i\Omega_{jk}$, where $\Omega$ is defined as
\begin{align}
\Omega = \begin{bmatrix}
0 & I_{N}\\
-I_{N} & 0
\end{bmatrix}
\end{align}
and $I_{N}$ is the $N\times N$ identity matrix.

By definition, the characteristic function of a Gaussian state
$\hat{\rho}$ is Gaussian\cite{Weedbrook2012}:
\begin{align}
\chi_{\hat{\rho}}(\boldsymbol{\mathrm{\xi}}) &\equiv \textrm{Tr}\big{[} \hat{\rho}\exp[i\boldsymbol{\hat{\mathrm{x}}}^{T}\Omega \boldsymbol{\mathrm{\xi}} ]\big{]}
\nonumber\\
&= \exp\Big{[} -\frac{1}{2}\boldsymbol{\mathrm{\xi}}^{T} ( \Omega V \Omega^{T}  ) \boldsymbol{\mathrm{\xi}} - i(\Omega\boldsymbol{\bar{\mathrm{x}}})^{T}\boldsymbol{\mathrm{\xi}} \Big{]},
\end{align}
where $\boldsymbol{\bar{\mathrm{x}}}$ and $V$ are the first and the second moments of the quadrature operator $\boldsymbol{\hat{\mathrm{x}}}$. A Gaussian state is thus fully characterized by its first two moments and one can write $\hat{\rho} = \hat{\rho}_{\textrm{G}}(\boldsymbol{\bar{\mathrm{x}}},V)$.

A Gaussian channel $\mathcal{N}$ is a completely positive and trace preserving map (a CPTP map)\cite{Choi1975} that maps a Gaussian state $\hat{\rho}_{\textrm{G}}(\boldsymbol{\bar{\mathrm{x}}},V)$ to another Gaussian state $\hat{\rho}_{\textrm{G}}(\boldsymbol{\bar{\mathrm{x}}}',V')$. A Gaussian channel $\mathcal{N}$ is fully characterized by its action on Gaussian states,
\begin{align}
\boldsymbol{\bar{\mathrm{x}}}' &= T\boldsymbol{\bar{\mathrm{x}}} + \boldsymbol{\mathrm{d}},
\nonumber\\
V' &= TVT^{T} + \bar{N},
\end{align}
i.e., by $(T,\bar{N},\boldsymbol{\mathrm{d}})$. A thermal-loss channel $\mathcal{N}[\eta,\nth]$ is a single-mode Gaussian channel that has $T = \sqrt{\eta}I_{2}$, $\bar{N} = (1-\eta) (\nth+\frac{1}{2})I_{2}$, and $\boldsymbol{\mathrm{d}} = \boldsymbol{\mathrm{0}}$ where $\eta\in[0,1]$ and $\nth\ge 0$. Thermal-loss channels are a good model for realistic optical and microwave communication channels. A thermal-loss channel with $\nth=0$ is called a bosonic pure-loss channel.

Other single-mode Gaussian channels include additive Gaussian noise channels and amplifier channels. An additive Gaussian noise channel $\mathcal{N}_{\textrm{B}_{2}}[\sigma]$ is characterized by $T = I_{2}$, $\bar{N} = \sigma^{2}I_{2}$, and $\boldsymbol{\mathrm{d}}=\boldsymbol{\mathrm{0}}$, and is also called a Gaussian random displacement channel. An amplifier channel $\mathcal{A}[G,\nth]$ is characterized by $T = \sqrt{G}I_{2}$, $\bar{N} = (G-1)(\nth+\frac{1}{2})I_{2}$, and $\boldsymbol{\mathrm{d}}=\boldsymbol{\mathrm{0}}$ where $G\ge 1$. An amplifier channel is called a quantum-limited amplification channel if $\nth=0$ and a noisy amplification channel if $\nth >0$ (see Sec. V of Ref. \citen{Weedbrook2012} for more details).

\subsection{Gaussian Fourier transformation} We define the $N$-mode Gaussian Fourier transformation $\hat{U}_{\textrm{GFT}}^{(N)}$ as a Gaussian operation that transforms the annihilation operators by a discrete Fourier transformation:
\begin{align}
\big{(} \hat{U}_{\textrm{GFT}}^{(N)} \big{)}^{\dagger} \hat{a}_{j} \hat{U}_{\textrm{GFT}}^{(N)} &= \frac{1}{\sqrt{N}}\sum_{k=1}^{N}e^{i\frac{2\pi}{N}(j-1)(k-1)}\hat{a}_{k}.
\end{align}
The $N$-mode Gaussian Fourier transformation can also be understood as a Gaussian (unitary) channel that is characterized by
\begin{align}
T = \begin{bmatrix}
R(0) & R(0) & \cdots & R(0)\\
R(0) & R(\frac{2\pi}{N}) & \cdots & R(\frac{2\pi}{N}(N-1))\\
\vdots & \vdots & \ddots & \vdots \\
R(0) & R(\frac{2\pi}{N}(N-1)) & \cdots & R(\frac{2\pi}{N}(N-1)^{2})
\end{bmatrix},
\end{align}
\textcolor{red}{$\bar{N}=0$}, and $\boldsymbol{\mathrm{d}}=0$, where
\begin{align}
R(\theta)\equiv \begin{bmatrix}
\cos\theta & -\sin\theta\\
\sin\theta & \cos\theta
\end{bmatrix}.
\end{align}
Since $T$ is an orthogonal matrix (i.e., $TT^{T} = T^{T} T = I_{2N}$), the $N$-mode Gaussian Fourier transformation is a passive linear optical operation that does not require squeezing.

\subsection{Proofs of Theorems \ref{theorem:Quantum capacity thermal loss} and \ref{theorem:two-way quantum capacity thermal loss} } Let $\mathcal{U}_{\textrm{GFT}}^{(N)}(\hat{\rho})\equiv \hat{U}_{\textrm{GFT}}^{(N)} \hat{\rho} (\hat{U}_{\textrm{GFT}}^{(N)})^{\dagger}$ be the unitary quantum channel associated with the $N$-mode Gaussian Fourier transformation. Then, $\mathcal{U}_{\textrm{GFT}}^{(N)}$ commutes with the tensor product of thermal-loss channels, i.e.,
\begin{align}
\mathcal{U}_{\textrm{GFT}}^{(N)} \mathcal{N}[\eta,\nth]^{\otimes N} = \mathcal{N}[\eta,\nth]^{\otimes N} \mathcal{U}_{\textrm{GFT}}^{(N)}. \label{method_eq:commutation between GFT and thermal loss}
\end{align}
This is a direct consequence of the fact that the $N$-mode Gaussian Fourier transformation is a passive linear optical operation with an orthogonal transformation matrix $T$. Now, recall that the correlated multi-mode thermal state $\hat{\mathcal{T}}(\boldsymbol{\mathrm{N}},\boldsymbol{\mathrm{n}})$ with $\boldsymbol{\mathrm{N}}=(N_{1},\cdots,N_{r})$ and $\boldsymbol{\mathrm{n}} = (\bar{n}_{1},\cdots, \bar{n}_{r})$ is defined as
\begin{align}
\hat{\mathcal{T}}(\boldsymbol{\mathrm{N}},\boldsymbol{\mathrm{n}}) = \mathcal{U}_{\textrm{GFT}}^{(N)} \Big{(}  \big{\lbrace}\hat{\tau}(\bar{n}_{1})\big{\rbrace}^{\otimes N_{1}}\otimes \cdots \otimes \big{\lbrace}\hat{\tau}(\bar{n}_{r})\big{\rbrace}^{\otimes N_{r}} \Big{)} . \label{method_eq:correlated multi-mode thermal state recalled}
\end{align}

Combining Eq. \eqref{method_eq:commutation between GFT and thermal loss} and Eq. \eqref{method_eq:correlated multi-mode thermal state recalled}, one can see that sending the correlated multi-mode thermal state $\hat{\mathcal{T}}(\boldsymbol{\mathrm{N}},\boldsymbol{\mathrm{n}})$ to the $N$ copies of thermal-loss channels is equivalent to sending a collection of thermal states $\big{\lbrace}\hat{\tau}(\bar{n}_{1})\big{\rbrace}^{\otimes N_{1}}\otimes \cdots \otimes \big{\lbrace}\hat{\tau}(\bar{n}_{r})\big{\rbrace}^{\otimes N_{r}}$ to the thermal-loss channels and then the receiver performing the Gaussian Fourier transformation. Since any local operations are assumed to be free, the achievable communication rates with the correlated multi-mode thermal state $\hat{\mathcal{T}}(\boldsymbol{\mathrm{N}},\boldsymbol{\mathrm{n}})$ is the same as the rates achievable with the collection of thermal states $\big{\lbrace}\hat{\tau}(\bar{n}_{1})\big{\rbrace}^{\otimes N_{1}}\otimes \cdots \otimes \big{\lbrace}\hat{\tau}(\bar{n}_{r})\big{\rbrace}^{\otimes N_{r}}$.

For quantum state transmission without any classical feedback assistance, the achievable rate is given by the coherent information. Since
\begin{align}
&I_{\textrm{c}} ( \mathcal{N}[\eta,\nth]^{\otimes N}, \hat{\mathcal{T}}(\boldsymbol{\mathrm{N}},\boldsymbol{\mathrm{n}}) )
\nonumber\\
&=I_{\textrm{c}}\Big{(} \mathcal{N}[\eta,\nth]^{\otimes N}, \big{\lbrace}\hat{\tau}(\bar{n}_{1})\big{\rbrace}^{\otimes N_{1}}\otimes \cdots \otimes \big{\lbrace}\hat{\tau}(\bar{n}_{r})\big{\rbrace}^{\otimes N_{r}} \Big{)}
\nonumber\\
&= \sum_{k=1}^{r} N_{k} I_{\textrm{c}}( \mathcal{N}[\eta,\nth] , \hat{\tau}(\bar{n}_{k}) ) , \label{eq:theorem 1 proof intermediate}
\end{align}
the correlated multi-mode thermal state $\hat{\mathcal{T}}(\boldsymbol{\mathrm{N}},\boldsymbol{\mathrm{n}})$ can achieve the quantum state transmission rate
\begin{align}
\frac{1}{N}\sum_{k=1}^{r} N_{k} I_{\textrm{c}}( \mathcal{N}[\eta,\nth] , \hat{\tau}(\bar{n}_{k}) )
\end{align}
per channel use. Specializing this to $\boldsymbol{\mathrm{N}} = (M,N-M)$ and $\boldsymbol{\mathrm{n}} = (\frac{N}{M}\bar{n},0)$, we get the rate
\begin{align}
\frac{M}{N}I_{\textrm{c}}(\mathcal{N}[\eta,\nth], \hat{\tau}\Big{(} \frac{N}{M}\bar{n} \Big{)} ) = xI_{\textrm{c}}( \mathcal{N}[\eta,\nth], \hat{\tau}\Big{(} \frac{\bar{n}}{x} \Big{)} )
\end{align}
as stated in Eq. \eqref{eq:thermal loss correlated multi mode thermal state coherent information} in Theorem \ref{theorem:Quantum capacity thermal loss}, where $x\equiv M/N$. Following the rest of the arguments given in Theorem \ref{theorem:Quantum capacity thermal loss}, the theorem follows.

Note that it might appear that the use of Gaussian Fourier transformation is not necessary because as shown in Eq. \eqref{eq:theorem 1 proof intermediate}, the coherent information of the correlated multi-mode thermal state $\hat{\mathcal{T}}(\boldsymbol{\mathrm{N}},\boldsymbol{\mathrm{n}})$ is the same as the coherent information of the uncorrelated multi-mode thermal state $\big{\lbrace}\hat{\tau}(\bar{n}_{1})\big{\rbrace}^{\otimes N_{1}}\otimes \cdots \otimes \big{\lbrace}\hat{\tau}(\bar{n}_{r})\big{\rbrace}^{\otimes N_{r}}$. It is nevertheless essential to use the Gaussian Fourier transformation because it uniformly spreads the excessive photons in \textcolor{red}{the} uncorrelated multi-mode thermal state across all modes such that the energy constraint is fulfilled (see also the discussion below Eq. \eqref{eq:Gaussian Fourier transforamtion}).   

Now consider the two-way quantum state transmission rate and the correlated multi-mode thermal states given in Theorem \ref{theorem:two-way quantum capacity thermal loss}, i.e., $\hat{\mathcal{T}}(\boldsymbol{\mathrm{N}},\boldsymbol{\mathrm{n}})$ with $\boldsymbol{\mathrm{N}}=(M,N-M)$ and $\boldsymbol{\mathrm{n}} = (\bar{n}_{1},\bar{n}_{2})$. As shown above, sending this state to a thermal-loss channel is equivalent to sending the collection of thermal states $\big{\lbrace} \hat{\tau}(\bar{n}_{1}) \big{\rbrace}^{\otimes M} \otimes \big{\lbrace} \hat{\tau}(\bar{n}_{2}) \big{\rbrace}^{\otimes N-M}$. Since classical feedback assistance is allowed, the reverse coherent information is also achievable in this case. By employing the forward strategy for the first $M$ modes and the backward strategy for the last $N-M$ modes, we can achieve the two-way quantum state transmission rate
\begin{align}
\frac{M}{N}I_{\textrm{c}}( \mathcal{N}[\eta,\nth] , \hat{\tau}(\bar{n}_{1}) ) + \frac{N-M}{N}I_{\textrm{rc}}( \mathcal{N}[\eta,\nth] , \hat{\tau}(\bar{n}_{2}) )
\end{align}
per channel use as stated in Eq. \eqref{eq:thermal loss correlated multi mode thermal state two-way rate} in Theorem \ref{theorem:two-way quantum capacity thermal loss}. Again, following the rest of the arguments in Theorem \ref{theorem:two-way quantum capacity thermal loss}, the theorem follows.

\subsection{Proof of Theorem \ref{theorem:private capacity thermal loss}} Recall that the classical-quantum state
\begin{align}
\hat{\sigma}(\bar{n}_{1},\bar{n}_{2}) \equiv \int d^{2}\alpha \frac{ e^{-|\alpha|^{2}/\bar{n}_{1}} }{\pi \bar{n}_{1}}  |\alpha\rangle\langle\alpha| \otimes \hat{D}(\alpha) \hat{\tau}(\bar{n}_{2}) \hat{D}^{\dagger}(\alpha)
\end{align}
with $\bar{n}_{1}+\bar{n}_{2} = \bar{n}$ was used to establish the recent lower bound in Eq. \eqref{eq:thermal loss private KS lower bound}. Similar to construction of correlated multi-mode thermal states in Theorem \ref{theorem:Quantum capacity thermal loss}, we construct the following classical-quantum state
\begin{align}
\hat{\Sigma}\equiv \Big{\lbrace} \hat{\sigma}\Big{(} \frac{N}{M} \bar{n} - \bar{n}_{2}^{\star} , \bar{n}_{2}^{\star} \Big{)} \Big{\rbrace}^{\otimes M} \otimes \big{\lbrace} \hat{\sigma}(0,0) \big{\rbrace}^{\otimes N-M}, \label{eq:classical-quantum state KN}
\end{align}
where $M\in \lbrace 1,\cdots ,N  \rbrace$. Then, by choosing
\begin{align}
\bar{n}_{2}^{\star} = \textrm{argmin}_{0\le \bar{n}_{2} \le \frac{N}{M}\bar{n} } I_{\textrm{c}}(\mathcal{N}[\eta,\nth] , \bar{n}_{2} ),
\end{align}
we can see that the achievable private communication rate of the state $\hat{\Sigma}$ is given by
\begin{align}
\frac{M}{N}f\Big{(} \eta,\nth, \frac{N}{M}\bar{n} \Big{)} = xf\Big{(} \eta,\nth, \frac{\bar{n}}{x} \Big{)} \label{method_eq:secure classical communication rate}
\end{align}
per channel use where $x=M/N\in (0,1]$ is a rational number. Since the set of rational numbers is a dense subset of the set of real numbers, the rate in Eq. \eqref{method_eq:secure classical communication rate} is achievable for any real number $x\in(0,1]$ and the theorem follows.

\end{methods}

\begin{addendum}
 \item[Data Availability] No data sets were generated during the current study.
\end{addendum}

\section*{References}
\vspace{0.4in}
\bibliographystyle{naturemag}
\bibliography{Superadditivity_v21_NC}


\begin{addendum}
 \item K.N. acknowledges support from the Korea Foundation for Advanced Studies. S.P. acknowledges support from the EPSRC via the `UK Quantum Communications HUB' (EP/M013472/1) and the European Union via the project `Continuous Variable
Quantum Communications' (CiViQ, no 820466). L.J. acknowledges support from the ARL-CDQI (W911NF-15-2-0067, W911NF-18-2-0237), ARO (W911NF-18-1-0020, W911NF-18-1-0212), ARO MURI (W911NF-16-1-0349 ), AFOSR MURI (FA9550-15-1-0015), NSF (EFMA-1640959), and the Packard Foundation (2013-39273). 
\item[Author Contributions] K.N. conceived this project. K.N. proved the theorems and generated the figures. K.N. and L.J. developed the convexity argument. S.P. developed and proved Theorem \ref{theorem:two-way quantum capacity thermal loss}. K.N., S.P., and L.J. wrote the manuscript.
 \item[Competing Interests] The authors declare that they have no
competing interests.
 \item[Correspondence] Correspondence and requests for materials
should be addressed to K.N. (email: kyungjoo.noh@yale.edu).
\end{addendum}

\end{multicols}
\newpage

\end{document}